\documentclass[%
 reprint,
 amsmath,amssymb,
 aps,
 pra,
]{revtex4-1}
\usepackage{subfigure}
\usepackage{graphicx}
\usepackage{dcolumn}
\usepackage{bm}

\begin{document}

\preprint{APS/123-QED}

\title{The role of near-field diffraction in photonic nanojet formation}

\author{Marouane Salhi}
\email{salhim@ornl.gov} 
\author{Philip G. Evans}%
 
\affiliation{%
 Quantum Information Science Group, Oak Ridge National Laboratory, Oak Ridge TN 37831, US\\
}%

\date{\today}

\begin{abstract}
In this paper, we show the photonic nanojet is the central bright maxima in focused near-field diffraction pattern. Using a simplified Huygens-Fresnel model, and using numerical simulations, we generate photonic nanojets from curved interfaces between media of different refractive indices and study the length, width, and peak intensity of the resultant photonic nanojet. We confirm that photonic nanojets are not sub-diffraction optical modes and are adequately described using diffraction theory. We also show how modifying the optogeometric environment can be used to tailor relevant features of the nanojets in general.
\end{abstract}

\pacs{Valid PACS appear here}

\maketitle

\section{Introduction}
\label{sec:intro}
The interest in nanophotonics, driven by promises of disruptive technologies across a wide variety of fields and supported by important research worldwide, continues to grow unabated. Nanophotonics typically refers to the generation, propagation, and interaction of photonic modes on the sub-wavelength scale, and includes plasmonics and near-field optics. The photonic nanojet (PNJ) is an atypical high-intensity beam and has attracted particular interest amongst the nanophotonics community. The PNJ concept was introduced by Chen \textit{et al.}~\cite{Chen2004} in 2004  using the finite difference time domain technique (FDTD) to study light scattering from a dielectric cylinder, confirmed by Li \textit{et al.}~\cite{Li2005} in 2005 again using FDTD and solving for scattered light from a sphere. The first experimental observation of PNJs were made by Heifetz \textit{et al.}~\cite{Heifetz2006} in 2006 using microwaves, followed by the direct imaging of PNJs produced by visible light scattering using confocal microscopy by Ferrand \textit{et al.}~\cite{Ferrand2008} in 2008. Since then, a wide variety of other structures have been shown to produce PNJs numerically ~\cite{Liu2012,Liu2013,Liu2015,McCloskey2015,Geints2015,Matsui2014} and experimentally ~\cite{Ferrand2008, McCloskey2015,Yang2011,Kim2012}. PNJs have been proposed for a wide variety of applications that include sub-wavelength imaging and microscopy ~\cite{Lee2015}, providing order-of-magnitude enhancements in Raman spectroscopy ~\cite{Dantham2011}, data storage ~\cite{Kong2008}, single molecule sensing ~\cite{Wenger2010}, nanofabrication ~\cite{Kim2012a,Kallepalli2013,CONSTANTINESCU2015} and even medicine ~\cite{Astratov2010}. In the work cited above, PNJ formation requires specific conditions, namely, the scattering object --- typically a sphere or cylinder --- must be of characteristic dimension \textit{R}, where $R \sim \lambda$, and having refractive index $\leq 2$ in vacuum \cite{Heifetz2009}. Under these conditions, neither the static field assumptions nor geometric optics approaches can fully describe the resultant electromagnetic field, and as such, Maxwell's equations were solved directly. Because of the inherent difficulties in studying nanophotonics and PNJs analytically, it is of no surprise that computational methods --- in particular the finite difference time domain (FDTD) approach --- have become \textit{de facto} tools for workers in the field to explore PNJs. In parallel with the computational and experimental efforts, several approaches have been proposed to analytically reproduce the PNJ using the uniform caustic asymptotic method ~\cite{Kofler2006}, Mie theory ~\cite{Lecler2005,Itagi2005}, and extended Mie theory ~\cite{Devilez2008,Geints2011}. However, formation of PNJs from non-spherical micro-optics cannot be described using Mie theory, leading to reliance on FDTD and other simulation tools to study PNJ formation in arbitrary geometries. This has left the study of PNJs lacking from a fundamental perspective, namely, a full understanding of how PNJs are formed and under what specific optogeometric conditions.

In this paper we show that photonic nanojets are accurately described using diffraction theory, and are essentially the result of focused beams in the near-field. Using a simplified Huygens-Fresnel approach, we demonstrate that near-field diffraction theory accurately describes both the length and peak position of the PNJ --- being identical to the central diffraction maxima in the near-field --- and show how it compares favorably to FDTD simulations of the same model. We clarify claims that the PNJ displays sub-diffraction properties from the standpoint of near-field diffraction, and note that previous works \cite{Linfoot1956, Bachynski1957, Evans1969} regarding image formation in optical instruments predict PNJ-like beam formation. Finally, we perform parametric studies of the optogeometric environment and demonstrate how features of the resultant PNJ can be tailored for a variety of applications of interest.

\section{Near-field diffraction}
\label{nearfielddiffraction}
Due to the small length scales involved, diffraction plays a significant role in nanophotonics, and while diffraction theory itself is well understood, the approximations and assumptions applied to simplify integrals in the diffraction calculations do not hold at the nanoscale. As a result, diffraction at the nanoscale, and generally in the near-field, is not as widely studied as far field diffraction. Consequently, light-matter interactions at the nanoscale are often treated as scattering problems, and while providing qualitative results, may fail to reveal the richness of the problem at hand.

Fraunhofer diffraction theory, providing an accurate account of geometric optics, results from several mathematical simplifications to the integral term in Kirchoff's diffraction theory. Specifically, the distance from the slit to an observation point $P$ is assumed to be much larger than the dimension of the slit itself, and consequently, only the first order terms in the integral expansion are retained. Thus, Fraunhofer diffraction accurately describes diffraction in the far-field, where small and rapidly decaying components of the diffracted wavefront close to the slit have no contribution. However, closer to the slit, in the observation region comparable to the slit dimension, one must retain second (and perhaps higher) order terms in the Kirchoff integral to account for near-field effects. Retaining the second order terms results in the Fresnel diffraction theory where, unlike Fraunhofer diffraction, curvature of the wavefront is required in order to account for the relative phase of interfering waves. By expanding to the second order, Fresnel diffraction retains the higher spatial frequency components of the diffracted wavefront. Consequently, assumptions correctly held in the far-field approach --- leading to concepts such as the well-known diffraction limit --- do not hold in the near-field. This is exploited by scanning probe optical microscopy - where features below the (far-field) diffraction limit are routinely imaged.

\begin{figure}[ht!]
\includegraphics[width=8.4cm]{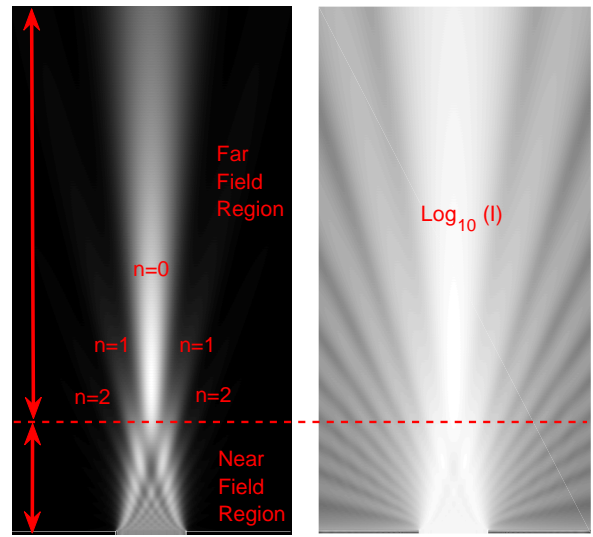}
\caption{Intensity field distribution for a near field diffraction pattern from an aperture of $5\mu m$ diameter and for an incident wavelength $\lambda = 0.55 \mu m$}
\label{fig1}
\end{figure}

\begin{figure}[ht!]
\includegraphics[width=8.4cm]{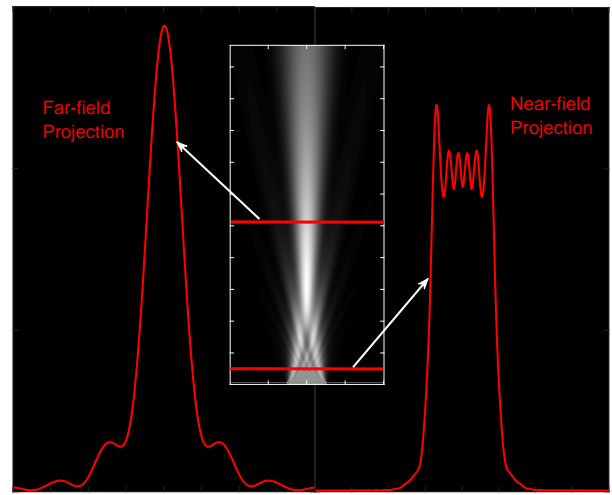}
\caption{Screen projection of Near Field Far Field light Diffraction}
\label{fig2}
\end{figure}

Yet solving diffraction problems in the near field are notoriously difficult from an analytical perspective. In a paper by Gillen and Guha ~\cite{Gillen2004}, the authors discussed several diffraction theory models and their regions of validity, aiming to obtain a complete description of the distribution of light immediately after the illuminated aperture. The complete Rayleigh-Sommerfeld diffraction model was found to accurately describe the intensity distribution pattern. Thankfully, using numerical tools such as those offered by the finite element method (FEM) and finite difference time domain (FDTD) techniques, one can reproduce their results by solving the full wave equations and without recurring to any approximations. Indeed, using FDTD simulations we obtain the diffracted light intensity distribution starting from the aperture and reaching to the "far-field region". As shown in Fig.\ref{fig1}, the diffraction pattern can be divided into two regions, near-field and far-field. The near field-region is well-defined by the pyramidal light distribution with an assembly of bright and dark fringes that has its base at the aperture. Starting from the base with many small bright fringes and ending at the summit with a unique large bright fringe that continues to the far-field region. Moreover, in Fig.\ref{fig2} we have extracted the field intensity in both regions. The far field intensity projection is what one would obtain using the Fraunhofer diffraction equation. It is evident that the near-field region contains the high spatial frequency modes which provide the richness to near-field optics.


\section{Focusing light in the near-field}
\label{focusing}
\begin{figure}[ht!]
\includegraphics[width=8.4cm]{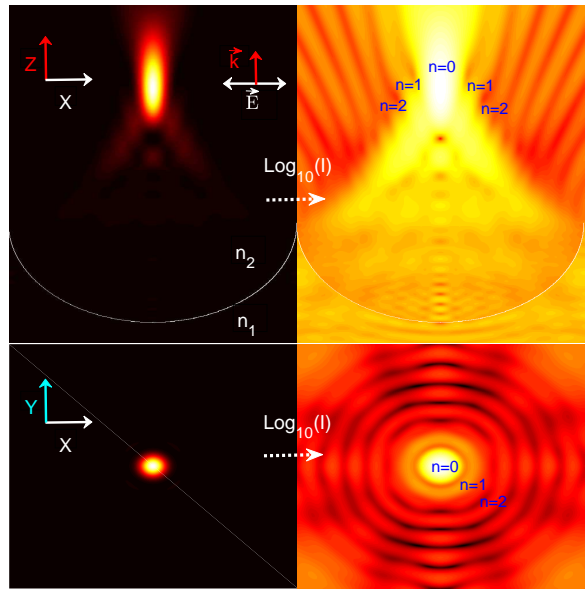}
\caption{Intensity field distribution of the PNJ obtained from an incident wavelength $\lambda=0.55\mu$m on circular interface of $5\mu$m diameter between two media of refractive indices $n_1=1$ and $n_2=1.5$ (left)  and the corresponding logarithmic function $\log(I)$ (right). In the lower right figure the Airy rings are evident.}
\label{fig3}
\end{figure}

\begin{figure}[ht!]
\includegraphics[width=6cm]{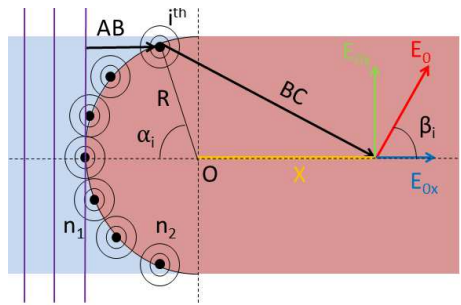}
\caption{Diffraction model system used to construct a PNJ along the axis of symmetry.}
\label{fig4}
\end{figure}

Historically, the interest in the intensity distribution near a focus started at the end of the nineteenth century ~\cite{VonLommel1885, Struve1886} to understand the theory of image formation in optical instruments. Since then, the subject has been addressed by many authors ~\cite{Joubin1892, Fabry1893, Julius1895, Zeeman1901, Sagnac1903, Debye1909, Reiche1909, Ignatowski1919, Fokker1923, Picht1930, Rubinowicz1938, Breuninger1939, Bouwkamp1940, ToraldodiFrancia1942}, generally involving different assumptions and approximations to overcome the complexity of the calculations. In 1949 Zernike and Nijboer ~\cite{Zernike1949} published realistic diagrams of the intensity distribution for images affected by spherical aberrations. In 1956, Linfoot and Wolf published a complementary work on the phase distribution near focus in an aberration free diffraction image ~\cite{Linfoot1956}. In the same year an experimental examination of the intensity distribution near the focus of microwave lenses was published by Bachynski and Bekefi ~\cite{Bachynski1957}. The main purpose of all these efforts was to gain an insight into the light distribution in the diffraction image of an optical system. The non-uniformity of the phase and amplitude of the incident wave upon the optical system strongly suggests the implementation of physical optics for an accurate evaluation of the obtained image. This is the case when the optical system size is on the order of one hundred wavelengths, thus the image is mainly determined by diffraction. The observation of PNJ satisfies this condition which suggests the use of diffraction to investigate its formation. Indeed when examining some of the isophote diagrams of the near field light intensity distribution constructed using diffraction theory and published more than sixty years ago ~\cite{Linfoot1956, Bachynski1957,Evans1969}, we can see what appears to be a PNJ. The lack of color-map figures showing the relative intensity of the focused beam may have played a role in hiding this important optical effect until it was introduced in 2004 ~\cite{Chen2004}. In fact, by examining closely the FDTD intensity field distribution of the PNJ emerging from a micro-scale hemispherical interface between two media and by applying the decimal logarithmic function to the intensity distribution as shown in Fig. \ref{fig3} to enhance the weak patterns, we can recognize the intensity pattern resulting form light diffraction from an aperture. Clearly, this pattern is suggesting that the PNJ is the result of light diffraction triggered by the circular interface of the dielectric medium. We consider the optical system illustrated in Fig. \ref{fig4}, consisting of a hemispherical dielectric interface between two media of refractive indices $n_1$ and $n_2$ and of radius $R$. A plane wave of wavelength $\lambda$ is incident upon the interface which acts as a diffracting slit.

In the typical treatment of diffraction using the Huygens-Fresnel principle, secondary sources of wavelets located within a slit in an opaque screen are assumed to be radiating in phase, i.e., the secondary sources are all equidistant from the primary source. However, this is not the case here where, due to the action of a plane wave meeting a curved surface, each secondary source on the lens surface will radiate with a relative phase shift with respect to others. Let us define an arbitrary point on the wave-front at $x = -R$ as A, and a corresponding point on the interface as $B$. We assume a secondary source of wavelets $i$ exists at $B$. The line $\overrightarrow{AB}$, parallel to the $x$-axis defines the distance from the primary wave front to the interface in vacuum, and is given by $R(1-cos(\alpha))$ where $\alpha$ is the polar angle defining the position of the secondary source $i$ . The relative phase shift of the secondary wavelets emitted from $B$ with respect to the secondary source located at $\alpha = 0$ is therefore
\begin{equation}
    \Delta \phi_{1} =\frac{2\pi n_1}{\lambda}R(1-cos(\alpha_i)).
\end{equation}
Let us then define a second line $\overrightarrow{BC}$ from the secondary source $B$ to an arbitrary point $C$ along the $x$-axis. The length of $\overrightarrow{BC}$ is $\sqrt{R^2+X^2+2 R X\cos{\alpha_i}}$. The relative phase shift experienced at $C$ due to the secondary source $i$ at $B$ is 
therefore

\begin{equation}
    \Delta \phi_{2} =\frac{2\pi n_2}{\lambda}\sqrt{R^2+X^2+2RX\cos{\alpha_i}}.
\end{equation}

Now the total phase shift at $C$ is the sum of $\Delta\phi_1+\Delta\phi_2$. We wish to evaluate the total electric field at arbitrary points along the $x$-axis. Due to the symmetry of our lens geometry, the superposition of two incident fields corresponding to the polar angles $\alpha_i$ and $- \alpha_i$ will null the $y$-components $E_{0y}$. Thus only the $x$-component of the electric field will be considered in evaluating the total field along the $x$-axis, $E_{0x} = E_0 \cos\beta_i$. Where $\beta_i$ is defined as

\begin{equation}
\beta_i = \arcsin\Big(\frac{BC^2+X^2-R^2}{2XBC}\Big).
\end{equation}

Integration over all $\alpha$ gives the total field at $C$ due to all secondary sources on the surface of the lens.

\begin{equation}\label{tot}
E_{tot}(X) =  E_0 \int_{-\frac{\pi}{2}}^\frac{\pi}{2} \! \cos{[\Delta\phi_1(\alpha_i)+\Delta\phi_2(X,\alpha_i)]} \cos{\beta_i} \, \mathrm{d}\alpha_i.
\end{equation}
 
\begin{figure}[ht!]
\includegraphics[width=7.5cm]{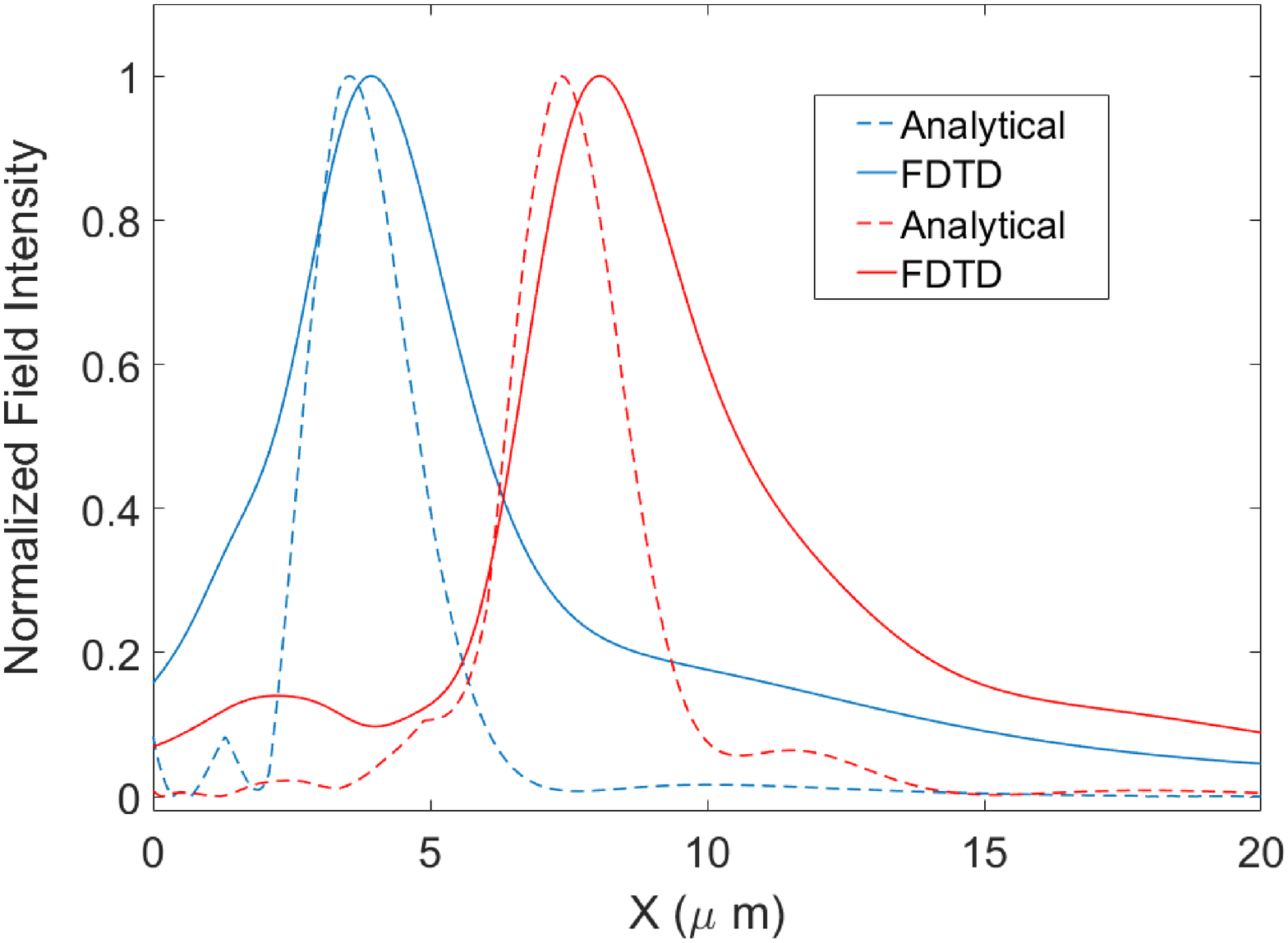}
\caption{PNJ normalized field intensity profile (envelope) for an incident wavelength $\lambda= 2\mu m$, refractive indices $(n_1=1,n_2=2)$ and a radius $ R =5 \mu m$ (blue), $ R =10 \mu m$ (red). The solid lines are obtained from the numerical simulations (FDTD in 2D) while the dashed lines are constructed using the derived expression of the total field along the X-axis (Eq. \ref{tot}).}
\label{fig5}
\end{figure}

Figure \ref{fig5} shows the normalized field intensity for an incident wavelength of 2 $\mu m$, $R$ = 5 and 10 $\mu m$, and $n_{1}$ = 1, $n_{2}$ = 2. FDTD simulations of the same geometry using the same parameters are performed with the corresponding results overlayed with the computational solution from the analytic diffraction approach. Both the FDTD technique and the analytic solution illustrate the common features of a PNJ --- namely the length, and the peak position --- with very good overlap between the analytical and numerical plots, where slight discrepancies are caused by our toy model not accounting for reflections at the optical interfaces, in addition to the finite meshing size used in the FDTD simulations. Nevertheless, using our simple diffraction model we have showed that we can reconstruct the salient features of a PNJ formed by a simple dielectric micro-optic lens.

\section{Dependence of the photonic nanojet on the optogeometric environment}
In this section, we use FDTD to study the dependence of PNJ width, length, and peak intensity on the refractive index $n_2$ of the dielectric medium, the incident wavelength $\lambda$ and the curvature ratio $\frac{R_1}{R_2}$ of the interface, as illustrated in Fig. \ref{fig8}. The PNJ length and width are defined by the FWHM of the intensity distribution. We present three sets of FDTD simulations showing the field intensity distributions of the PNJ: (1) where the refractive index $n_2$ of the dielectric medium is varied, (2) where the wavelength of incident plane wave is varied, and (3) where the curvature ratio $\frac{R_1}{R_2}$ is varied.

\subsection{PNJ dependence on the refractive index $n_2$}
\begin{figure}[ht!]
\includegraphics[width=8.4cm]{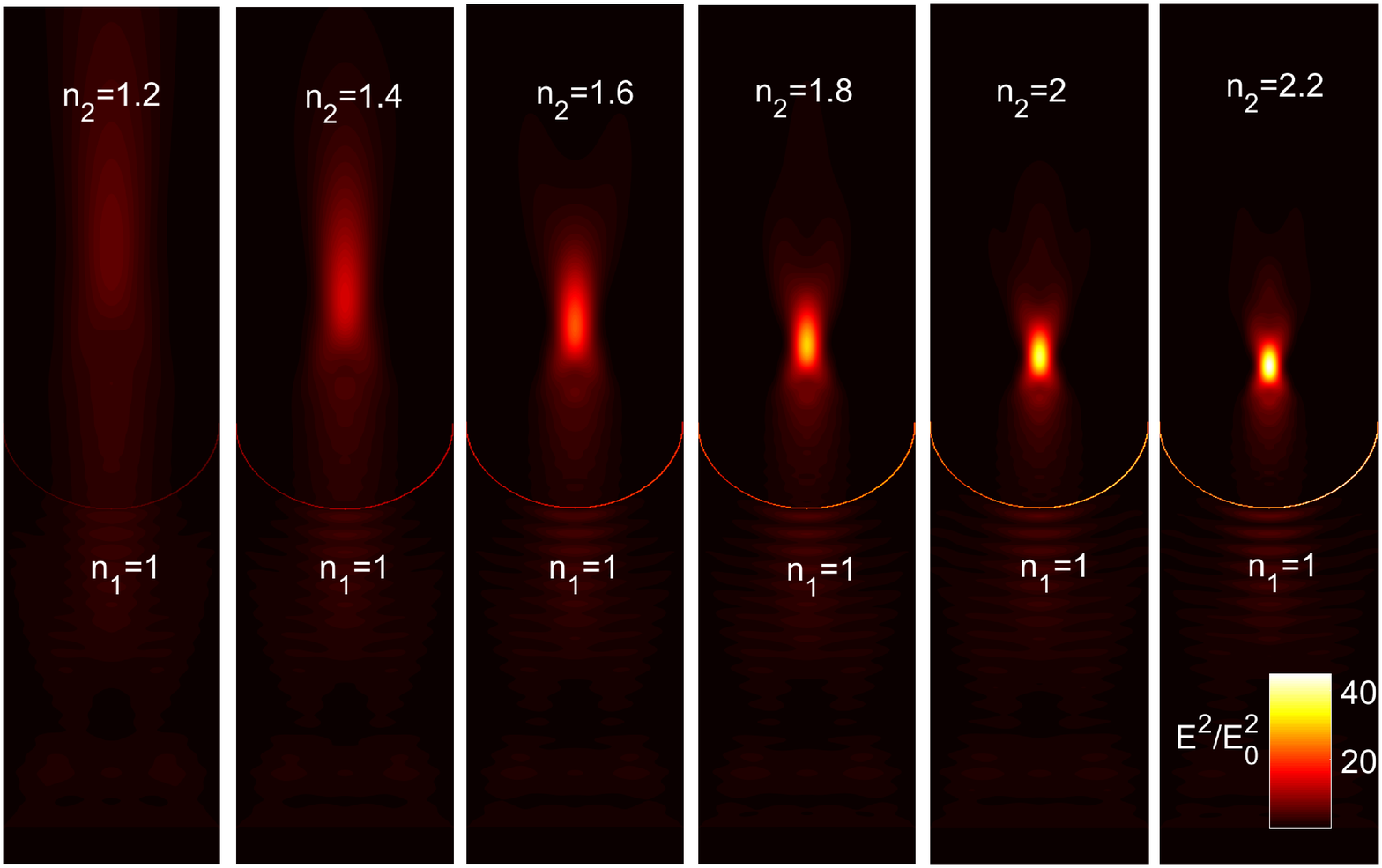}
\vskip\baselineskip
\includegraphics[width=8.4cm]{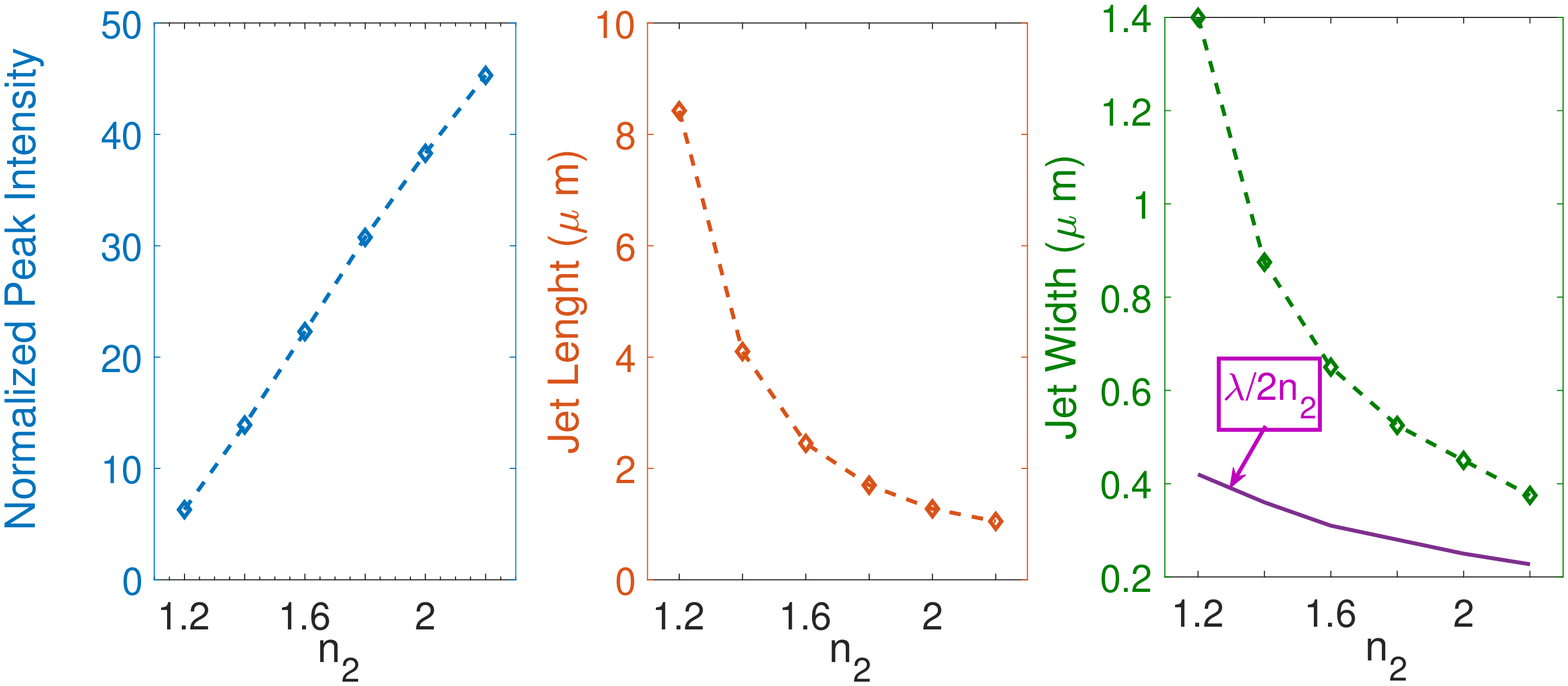}
\caption{PNJ intensity field intensity distribution for an incident wavelength $\lambda= 1\mu m$, a variation of the refractive index $n_2$ in $[1.2-2.2]$, a fixed refractive index $n_1=1$ with a fixed circular interface radius $ R =2.5 \mu m$.}
\label{fig6}
\end{figure}

The first set of simulations involves variation of $n_2$ over the range {1.2 - 2.2} for a fixed hemispherical interface of radius $R = 2.5 \mu m$, with fixed incident wavelength $\lambda = 1 \mu m$ and background index $n_1 = 1$. Results of the simulations are shown in Fig. \ref{fig6}. As $n_2$ is increased from 1.2 to 2.2, we show the PNJ the peak intensity increases by an order of magnitude. Both the jet length and width decreases, as the position of the jet's peak intensity moves closer to the interface, demonstrating a focusing effect. However, as we show in the lower right panel of Fig. \ref{fig6}, the width of the jet does not decrease below $\lambda / 2n_2$, the far-field diffraction limit.

\subsection{PNJ dependence on the incident wavelength $\lambda$}
\begin{figure}[ht!]
\includegraphics[width=8.4cm]{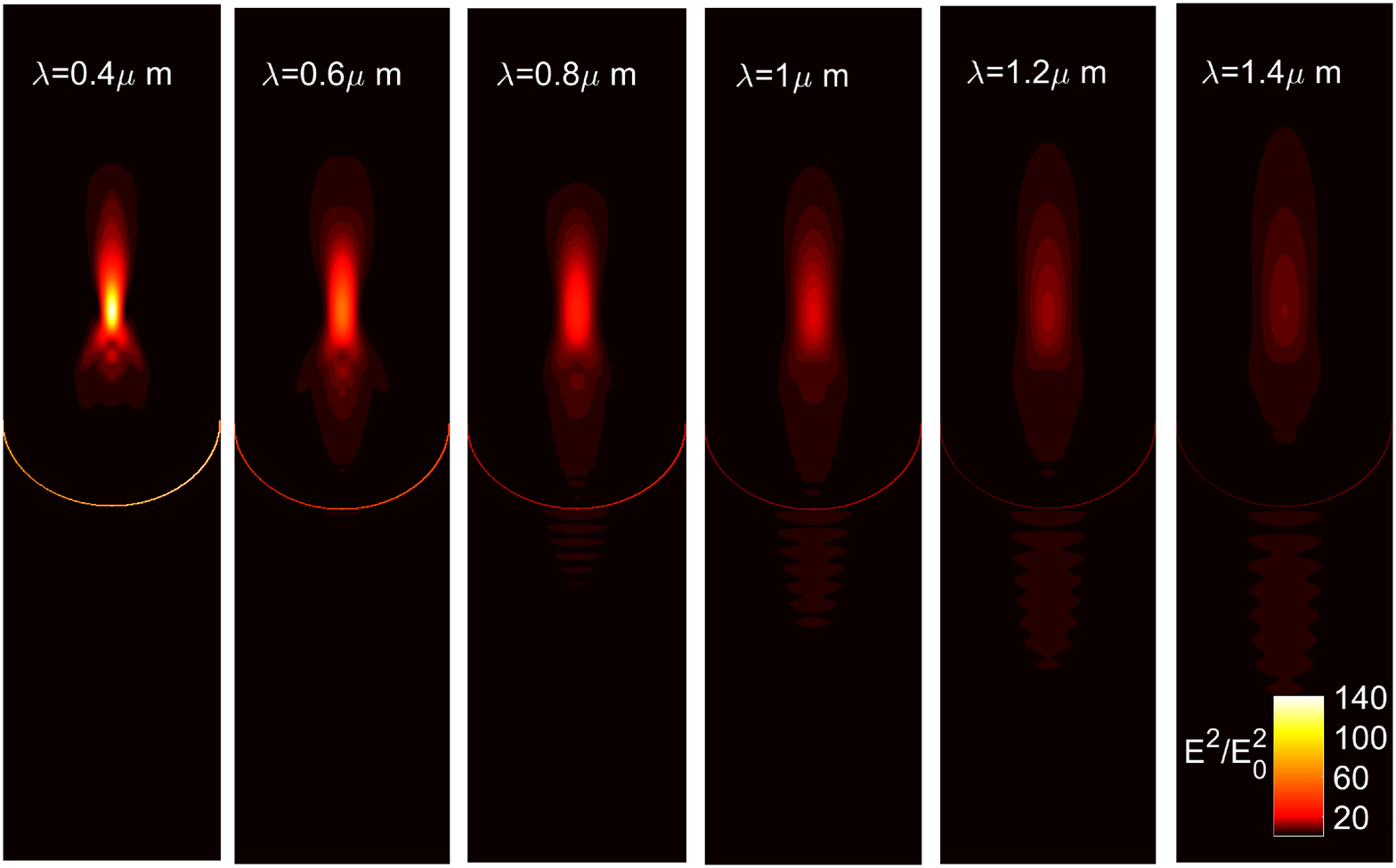}
\vskip\baselineskip
\includegraphics[width=8.4cm]{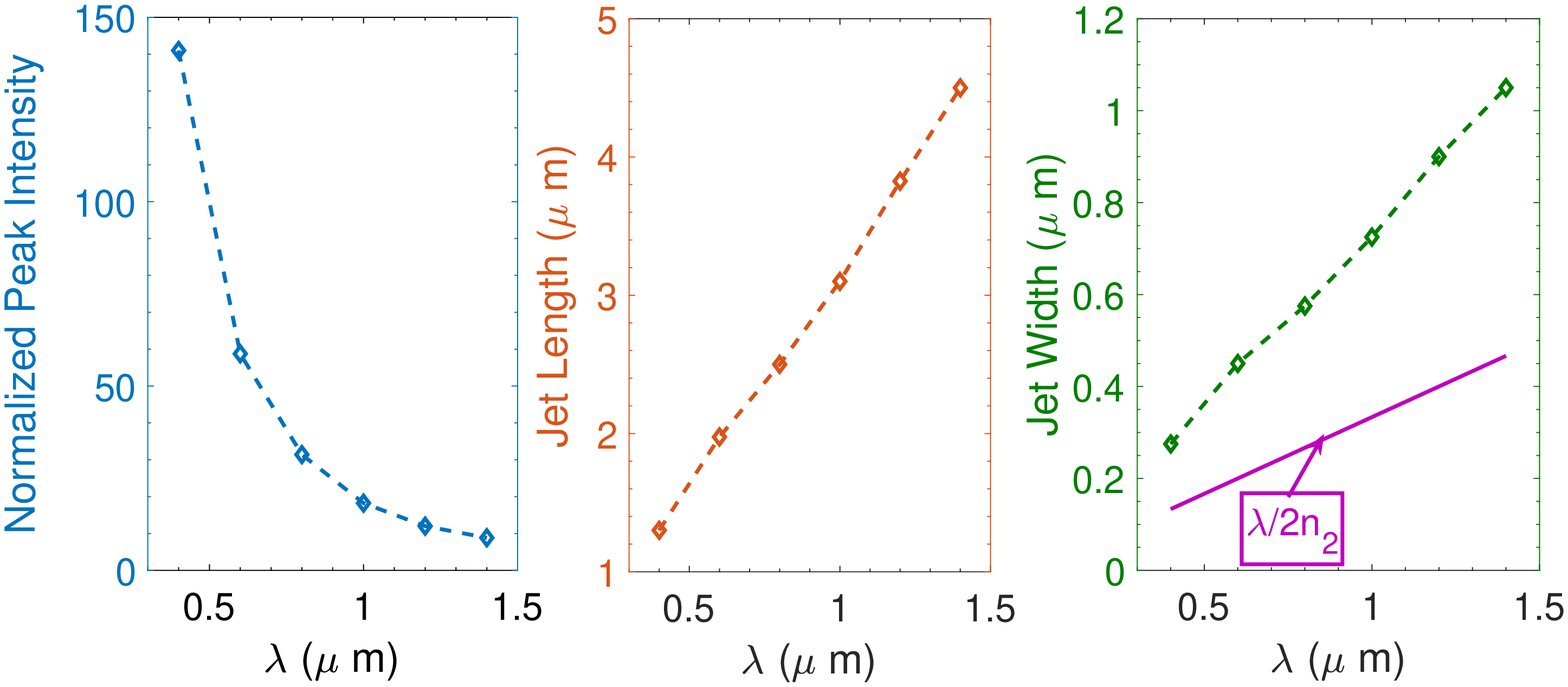}
\caption{PNJ intensity field intensity distribution for a variation of the incident wavelength $\lambda$ in $[0.4-1.4]\mu$m, fixed refractive indices $n_1=1$ and $n_2=1.5$, with a fixed circular interface of radius $ R =2.5 \mu m$.}
\label{fig7}
\end{figure}

The second set of simulations involves variation of $\lambda$ over the range {0.4 - 1.4 $\mu m$} for a fixed hemispherical interface of radius $R = 2.5 \mu m$, with fixed $n_2 = 1.5$ and background index $n_1 = 1$. Results of the simulations are shown in Fig. \ref{fig7}. As $\lambda$ is increased from 0.4 $\mu m$ to 1.4 $\mu m$, we show the PNJ the peak intensity decreases by almost two orders of magnitude. Both the jet length and width display a linear increase with increasing wavelength. Again, as we show in the lower right panel of Fig. \ref{fig7}, the width of the jet does not decrease below $\lambda / 2n_2$, the far-field diffraction limit.

\subsection{PNJ dependence on the curvature ratio $\frac{R_1}{R_2}$}
\begin{figure}[ht!]
\includegraphics[width=8.4cm]{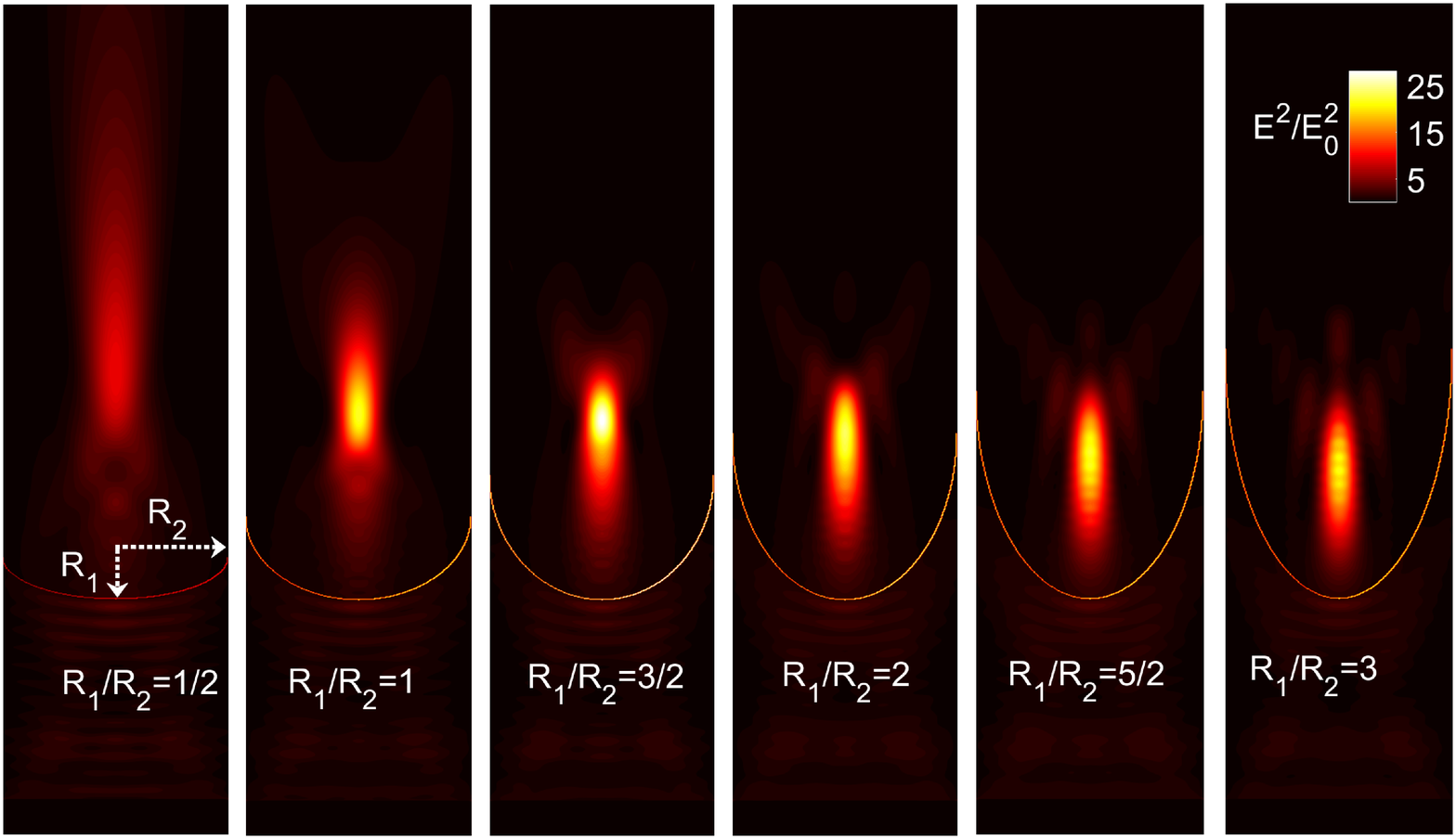}
\vskip\baselineskip
\includegraphics[width=8.4cm]{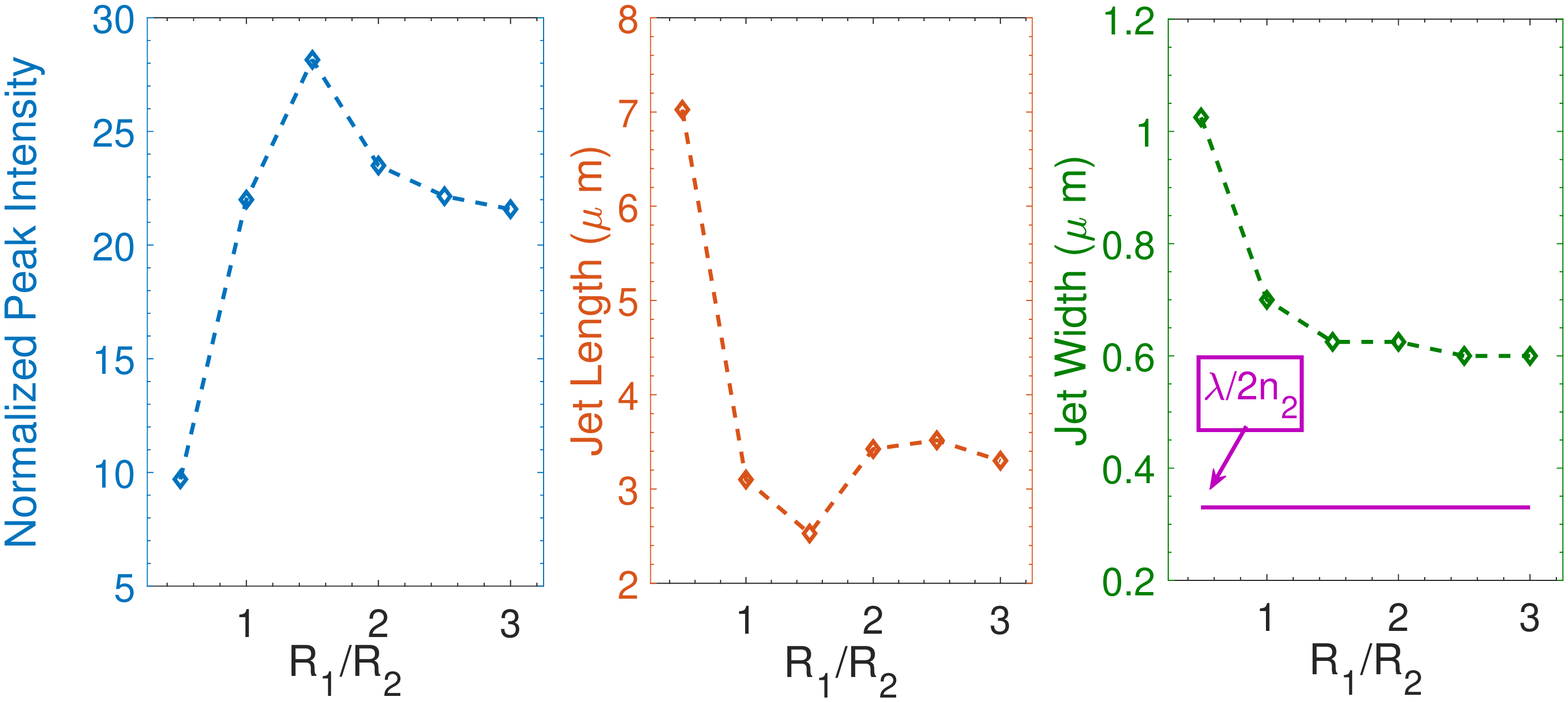}
\caption{PNJ intensity field intensity distribution for an incident wavelength $\lambda= 1\mu m$, fixed refractive indices $n_1=1$ and $n_2=1.5$, with a variation of the curvature defined as the ratio $\frac{R_1}{R_2}$}
\label{fig8}
\end{figure}

The final set of simulations involves variation of the curvature ratio $\frac{R_1}{R_2}$, over the range {0.5 - 3}, of an ellipsoidal interface between the background of $n_1 = 1$ and fixed dielectric $n_2 = 1.5$. The incident wavelength is fixed at $\lambda = 1 \mu m$. $R_1$ corresponds to the axis parallel to the direction of propagation, whereas $R_2$ is perpendicular. Results of the simulations are shown in Fig. \ref{fig8}. Unlike in the previous two cases, there is not a clear consequence on varying the curvature of the interface. However, we note that at $\frac{R_1}{R_2} = 1.5$, we see a maxima in the peak intensity, and a corresponding minima in the jet length. As $\frac{R_1}{R_2}$ increases, there is an exponential decrease in the jet width, appearing to reach an asymptote of $\sim 0.6 \mu m$ while still not decreasing below $\lambda / 2n_2$.

\section{Conclusion}
\label{conclusion}
The photonic nanojet is a consequence of focused near field diffraction from a micron scale dielectric object. The PNJ corresponds to the central bright maxima of the diffraction pattern. A curved micro-scale interface between two media of different refractive indices operates as a diffracting slit and a focusing lens at the same time and causes the formation of the PNJ. Understanding the mechanism behind the formation of the nanojet has eliminated a number of constraints believed to be necessary to observe the PNJ, such as the choice of a particular dielectric geometry, refractive index, dimensions and specific incident wavelength. As shown in the parametric study the main features of the PNJ (length, width and peak intensity) are essentially controlled with two focusing and de-focusing knobs, the refractive index $n_2$ and wavelength $\lambda$.

\section*{Funding}
This research was supported by an appointment to the Intelligence Community Postdoctoral Research Fellowship Program at Oak Ridge National Laboratory, administered by Oak Ridge Institute for Science and Education through an interagency agreement between the U.S. Department of Energy and the Office of the Director of National Intelligence.
\section*{Acknowledgments}
This manuscript has been authored by UT-Battelle, LLC under Contract No. DE-AC05-00OR22725 with the U.S. Department of Energy. The United States Government retains and the publisher, by accepting the article for publication, acknowledges that the United States Government retains a non-exclusive, paid-up, irrevocable, worldwide license to publish or reproduce the published form of this manuscript, or allow others to do so, for United States Government purposes. The Department of Energy will provide public access to these results of federally sponsored research in accordance with the DOE Public Access Plan (http://energy.gov/downloads/doe-public-access-plan).
\bibliography{PNJ}
\end{document}